\documentclass[pre,preprint,aps,eqsecnum]{revtex4}

\usepackage{graphicx}

\begin{document}

\title{Thermal Effects in the Shear-Transformation-Zone Theory of Amorphous
Plasticity: Comparisons to Metallic Glass Data}

\author{M. L. Falk} \affiliation{Department of Materials Science and
Engineering,  University of Michigan,  Ann Arbor, MI 48109-2136}

\author{J. S. Langer} \author{L. Pechenik}

\affiliation{Department of Physics, University of California, Santa
Barbara, CA  93106-9530  USA}

\date{November, 2003}

\begin{abstract}
We extend our earlier shear-transformation-zone (STZ) theory of
amorphous plasticity to include the effects of thermally assisted
molecular rearrangements.  This version of our theory is a substantial
revision and generalization of conventional theories of flow in
noncrystalline solids.  As in our earlier work, it predicts a dynamic
transition between jammed and flowing states at a yield stress.  Below
that yield stress, it now describes thermally assisted creep.  We show
that this theory accounts for the experimentally observed strain-rate
dependence of the viscosity of metallic glasses, and that it also
captures many of the details of the transient stress-strain behavior
of those materials during loading.  In particular, it explains the
apparent onset of superplasticity at sufficiently high stress as a
transition between creep at low stresses and plastic flow near the
yield stress. We also argue that there are internal inconsistencies in
the conventional theories of these deformation processes, and suggest
ways in which further experimentation as well as theoretical analysis
may lead to better understanding of a broad range of nonequilibrium
phenomena.
\end{abstract}
\maketitle

\section{Introduction}

In the first paper of this series\cite{STZdynsI}, we showed that
energetic constraints determine the principal ingredients of a
shear-transformation-zone (STZ) theory of amorphous plasticity.  That
analysis pertained strictly to the behavior of noncrystalline solids
well below their glass temperatures.  We turn our attention here to
the roles played by thermal fluctuations, specifically, to the ways in
which glassy materials make transitions from thermally activated creep
to viscoplastic flow near yield stresses.

The atomic mechanisms of plastic deformation are most often described as
arising from dislocation motion.  This picture breaks down in amorphous
solids in which the dislocation, being a lattice defect, ceases to provide a
useful description of the microstructural dynamics.  In this paper, we 
describe further progress in the shear-transformation-zone theory of amorphous
plasticity that we originally constructed to relate plastic deformation to
specific microstructural degrees of freedom in non-crystalline solids.
From its inception, our STZ picture has been based on the ideas of
Cohen, Turnbull, Spaepen, Argon and others
\cite{TURNBULL,SPAEPEN77,SPAEPEN81,ARGON}, who postulated that plastic
deformation in amorphous materials occurs at localized sites usually
called flow defects.  A number of computational studies
(e.g. \cite{SROLOVITZ,DENG}) have provided support for the idea that a
model based on localized defects can capture the dynamics of
deformation in such systems.  The basic premise of our version of the
STZ theory is that these defects must be dynamic entities that carry
orientational information; and our most striking conclusion is that,
once these orientational degrees of freedom are taken into account,
the system so described exhibits an exchange of dynamic stability
between jammed and flowing states at a stress that we identify as a
yield stress.

Earlier defect theories of deformation in glassy materials appear to
us to be incomplete in important respects and, in some cases, to
contain physically unrealistic assumptions.  These theories generally
start by assuming that the plastic strain rate $\dot\epsilon^{pl}$ is
the product of the density of defects $n$ times an Eyring rate
factor\cite{EYRING}:
\begin{equation}
\label{epsdotintro}
\dot\epsilon^{pl}= 2\,n\,\nu\,\exp\left(-{\Delta G\over
k_B\,T}\right)\,\sinh\left({\Omega\,s\over 2\,k_B\,T}\right),
\end{equation}
where $\nu$ is a molecular vibration frequency, $\Delta G$ is an
activation barrier, $k_B$ is Boltzmann's constant, $T$ is the
temperature, $\Omega$ is an atomic volume, and $s$ is the deviatoric
stress (i.e. the shear stress).  The authors of these theories then
attempt to describe the deformation dynamics by postulating equations
of motion for $n$.  The most common choice for such an equation of
motion has the form
\begin{equation}
\label{ndotintro}
\dot n = -k_r\,n\,(n - n_{eq})+ P(\dot\epsilon^{pl}).
\end{equation}
Here, $k_r$ is a thermally activated rate
factor, $n_{eq}$ is the thermal equilibrium density of flow defects in
the absence of external driving forces, and $P$ is a production rate
that vanishes when the strain rate is zero.  
An important example of the use of this equation is the paper by De
Hey et al. \cite{DEHEY}. As we shall see, our STZ
theory has many features in common with Eq.(\ref{ndotintro}); indeed,
some form of each of the terms in this equation will appear here.

There are important differences, however.  Theories based on
Eqs. (\ref{epsdotintro}) and (\ref{ndotintro}) make no attempt to
describe what actually happens when a zone undergoes a shear
transformation.  Despite the fact that the Eyring formula in
Eq.(\ref{epsdotintro}) describes the balance between forward and
backward transitions in some kind of two-state system, the dynamics of
atomic scale structural rearrangements described by
Eq.(\ref{epsdotintro}) is decoupled from the population dynamics
described by Eq.(\ref{ndotintro}).  The latter equations, therefore,
implicitly assume that there exists some fast relaxation mechanism --
faster than any other rate introduced explicitly in the theory --
which causes zones instantaneously to lose their memory of prior
transformations.

Another major difference is in the choice of the production function
$P$.  In some theories, $P$ is chosen to be linearly proportional to
$\dot\epsilon^{pl}$, which is impossible because $P$ must be a
non-negative scalar while $\dot\epsilon^{pl}$ is a tensor that can
change sign.  Such theories generally are used only in cases where
$\dot\epsilon^{pl}$ is positive, which may mean that the authors
intend to use the magnitude of $\dot\epsilon^{pl}$ in more general
situations; but the latter convention also would be unsatisfactory
because $|\dot\epsilon^{pl}|$ is a nonanalytic function that is not
likely to arise from any first-principles analysis of molecular
mechanisms.  In our own earlier paper \cite{FL}, we tried using the
rate of plastic work, $s\,\dot\epsilon^{pl}$, in our analog of the
production term.  That function is a scalar with a satisfactory
physical interpretation but, as we remarked there, it also suffers
from a sign problem because it can be negative during unloading.  We
believe that we have solved this problem in \cite{STZdynsI}, and shall
make extensive use of the technique described there in what follows.

Spaepen \cite{SPAEPEN77,SPAEPEN81} has introduced an important
modification of the above ideas by postulating that the defect density
$n$ is directly determined by the excess free volume in the system,
$v_f$, {\it via} a relation of the form
\begin{equation}
\label{vfintro}
n \propto \exp\left(-{V^*\over v_f}\right),
\end{equation}
where $V^*$ is a molecular volume.  Spaepen's proposal is that the
production term $P$ in Eq.(\ref{ndotintro}) should be proportional to
the growth rate of $v_f$. More recently, Johnson et al.\cite{JOHNSON} have
proposed a dynamic free volume model that includes a phenomenological
parameter very roughly analogous to the yield stress that emerges from 
our STZ theory.  Unfortunately, Spaepen, Johnson and others
(e.g. \cite{DEHEY}) postulate an equation of the form $\dot v_f
\propto \dot\epsilon^{pl}$, which again violates symmetry or
analyticity requirements. In short, we believe that these theories
require a critical reformulation in order to enable a meaningful
atomic-scale analysis of amorphous plasticity.

In addition to reformulating earlier theories of amorphous plasticity,
one of our principal goals in this paper is to gain as simple as
possible an understanding of recent experimental results on plastic
flow in metallic glasses.  In particular, we refer to work by Kato et
al.\cite{KATO} on amorphous ${\rm Pd_{40}\,Ni_{10}\,Cu_{30}\,P_{20}}$,
and the results of Lu et al.\cite{LU}, who measured properties of bulk 
amorphous ${\rm Zr_{41.2}\,Ti_{13.8}\,Cu_{12.5}\,Ni_{10}\,Be_{22.5}}$ over a
remarkably wide range of strain rates and temperatures.  

For the sake of simplicity, we restrict ourselves throughout this paper to the 
quasilinear theory discussed in \cite{STZdynsI}. We also make  other 
assumptions which, as we shall indicate at several places in our presentation, 
raise fundamental issues regarding nonequilibrium states in deforming
solids.

Our order of presentation is as follows.  In Section II, we review the
elements of the low-temperature, quasilinear theory \cite{STZdynsI} in
a way that prepares concepts and notation for the discussion of
thermal relaxation in Section III.  Section IV is devoted to analysis
of these theoretical results and comparison with experiments.
Finally, in Section V, we return to various fundamental issues that
are raised earlier in the paper or are relevant to its conclusions.
In particular, we discuss how our original, fully nonlinear STZ theory
\cite{FL} may need to be invoked in order to make better contact with
atomic-scale mechanisms; we briefly address the important issue of
shear banding; and we discuss the question of departures from thermodynamic 
equilibria in driven systems -- a question that we shall not be able to avoid 
in the present analysis.

\section{Elements of the Low-Temperature STZ Theory}

In order that this account be reasonably self-contained, and to
provide some new perspectives about the discussion that follows, we
start by reviewing the basic elements of the low-temperature STZ
theory.

Throughout the following analysis, we take our original STZ picture
\cite{FL} more literally than perhaps is necessary.  We assume that,
instead of being structureless objects as in the earlier theories
described in the Introduction, the STZ's are two-state systems. In the
presence of a shear stress, they can deform by a finite amount in one
direction before becoming jammed and, when jammed in one direction,
they can transform in the opposite direction in response to a reversed
stress. Importantly, our STZ's are ephemeral; they are created and
annihilated during irreversible deformations of the material.

The literal interpretation to be used here requires that all STZ's
have approximately the same size and dynamic properties.  To visualize
such an STZ, think of a void in an elastic material, and place a small
group of molecules inside it in such a way that their average free
volume is somewhat larger than that for most other molecules in the
system.  The void has some degree of structural stability; it can
deform elastically but, because of the configuration of molecules on
its surface, it resists collapse.  Rearrangements of the molecules
that are caged within the void couple to its shape and, therefore, to
the stress field in the elastic medium in which the void is
embedded. This picture suggests that the system undergoes two distinct
kinds of irreversible events: volume conserving shear deformations,
i.e. the STZ transformations, and dilations or contractions in which
the STZ's are created or annihilated.  (Nothing in this picture rules
out the possibility that transient dilations occur during intermediate
stages of the shear transformations.)

For simplicity, as in \cite{STZdynsI,FL}, we consider a two-dimensional model,
and we subject this system only to pure shear deformations. As we
shall show at the beginning of Section IV, the properties of this
model are easily reinterpreted in terms appropriate to simple uniaxial
stress experiments in three dimensions, so long as we are willing to
assume that the system remains spatially uniform.
We further restrict ourselves to molecular materials in
contact with thermal reservoirs (as opposed to granular materials or
foams), so that we may assume that an ambient temperature determines
an underlying fluctuation rate.  At low temperatures, thermal
fluctuations provide an attempt frequency for stress-induced molecular
rearrangements, but are too small to activate transitions over energy
barriers in the absence of external driving forces. This attempt
frequency will never appear explicitly in our analysis but, rather, is
embedded in the rate factors to be introduced below.

For present purposes, we need to consider only situations in which the
orientation of the principal axes of the  stress and strain tensors
remains fixed. That is, we do not consider situations in which a fully
tensorial version of the STZ theory will be necessary, as in the
necking calculations reported in \cite{ELP}.  Therefore, it is
sufficient to assume that the population of STZ's consists simply of
zones oriented along the two relevant principal axes of the stress
tensor.  Exactly the same equations as the ones we shall use here can
be derived starting from the assumption that the {\it a priori}
distribution of orientations of the zones is continuous and
symmetric.\cite{Fthesis,PECHENIK} Without loss of generality,
therefore, we let the deviatoric stress be diagonal along the $x$, $y$
axes; specifically, let $s_{xx}=-s_{yy}= s$ and $s_{xy}=0$. Then
choose the ``$+$" zones to be oriented (elongated) along the $x$ axis,
and the ``$-$" zones along the $y$ axis; and denote the population
density of zones oriented in the ``$+$/$-$'' directions by the symbol
$n_{\pm}$.

With these conventions, the plastic strain rate is:
\begin{equation}
\label{epsdot}
\dot\epsilon_{xx}^{pl}= -\dot\epsilon_{yy}^{pl}\equiv
\dot\epsilon^{pl}={\lambda\over\tau_0}\,\Bigl(R(-s)\,n_--R(s)\,n_+\Bigr).
\end{equation}
Here $\lambda$ is a material-specific parameter with the dimensions of
volume (or area in strictly two-dimensional models), which must be
roughly the same order of magnitude as the volume of an STZ, that is,
a few  cubic or square atomic spacings.  The remaining factor on the
right-hand side of Eq.(\ref{epsdot}) is the net rate per unit volume
at which STZ's transform from ``$-$'' to ``$+$'' orientations.
$R(s)/\tau_0$ and $R(-s)/\tau_0$ are the rates for ``$+$" to ``$-$''
and ``$-$" to ``$+$'' transitions respectively, where $\tau_0$ is the
time scale that characterizes the low-temperature plastic
response. For simplicity, we write these rates as explicit functions
of only the deviatoric stress $s$, although they depend implicitly on
the temperature and pressure and perhaps other quantities.

The equation of motion for the populations $n_{\pm}$ generally must be
a master equation of the form:
\begin{equation}
\label{ndot}
\tau_0\,\dot n_{\pm}=R(\mp s)\,n_{\mp}-R(\pm s)\,\,n_{\pm}
+\Gamma(s,...)\,\left({n_{\infty}\over 2}-n_{\pm}\right).
\end{equation}
The first two terms on the right-hand side are the stress-driven
transition rates introduced in the preceding paragraph.  There is no
analog of these terms in Eq.(\ref{ndotintro}).  They describe
volume-conserving, pure-shear deformations which preserve the total
population of the STZ's.  The last two terms in parentheses,
proportional to $\Gamma$, describe creation and annihilation of STZ's.
In the low-temperature theory, $\Gamma$ is nonzero only when the
plastic strain rate is nonzero; the molecular rearrangements required
for creating or annihilating STZ's cannot occur spontaneously, that
is, in the absence of external driving forces

The assumption in Eq.(\ref{ndot}) that the annihilation and creation
rates are both proportional to the same function $\Gamma$ has serious
implications in this theory. Among those implications is the
requirement that $n_{\infty}$ be a strain-rate independent constant.
Note that $n_{\infty}$ is the total density of zones generated in a
system that is undergoing steady plastic deformation.  It is not the
same as the quantity $n_{eq}$ introduced in Eq.(\ref{ndotintro}),
which is the equilibrium density at nonzero temperature and zero
strain rate, and ordinarily is said to go rapidly to zero as the
temperature decreases below the glass transition.  On the other hand,
$n_{\infty}$ is a property of low-temperature materials at non-zero
strain rates.

The form in which we have cast Eq.(\ref{ndot}) is consistent with a
fundamental assumption that we are making about the nature of our
low-temperature theory.  Specifically, we are assuming that the only 
relevant time scales at low 
temperatures are $\tau_0$ and the inverse of the strain rate.  This means 
that, under steady-state conditions at strain rates less than some value of 
order $\tau_0^{-1}$, the number of events in which the molecules rearrange 
themselves is not proportional to the time but to the strain.
That picture seems intuitively reasonable.  If the system requires a
certain number of STZ-like rearrangements in order to achieve some
deformation, then it should not matter (within limits) how fast that
deformation takes place.  The picture breaks down, of course, when
there are competing rearrangement mechanisms.  For example, we shall
see that the density of STZ's becomes strain-rate dependent when we
introduce thermal fluctuations, because such fluctuations will induce
rearrangements at a rate that is independent of the strain rate.  We
also expect that the picture may fail in polymeric glasses or
polycrystalline solids, where more complex components may introduce
extra length and time scales.

This simple dimensional argument, leading to a nonzero,
rate-independent value of $n_{\infty}$, already hints at a fundamental
difficulty in theories of the kind summarized by
Eqs.(\ref{epsdotintro}) and (\ref{ndotintro}).  These theories have no
sensible low-temperature limit because both the strain rate in
Eq.(\ref{epsdotintro}) and the rate factor $k_r$ in
Eq.(\ref{ndotintro}) vanish rapidly as $T\to 0$.  Yet, even at
temperatures so low that thermal fluctuations cannot cause molecular
rearrangements, such systems must deform irreversibly when sheared.

We shall use the energetic arguments introduced in \cite{STZdynsI} to
determine the factor $\Gamma$ in Eq.(\ref{ndot}), but first we must
discuss the state variables and specific forms for the transition
rates.  We define dimensionless state variables by writing
\begin{equation}
\label{vardef}
\Lambda \equiv {n_++n_-\over n_{\infty}},~~~~\Delta\equiv
{n_+-n_-\over n_{\infty}}.
\end{equation}
In a more general treatment \cite{SHEARLOC,ELP,PECHENIK}, $\Lambda$
remains a scalar density, but $\Delta$ becomes a traceless symmetric
tensor with the same transformation properties as the deviatoric
stress.  We also define:
\begin{equation}
\label{Tdef}
{\cal S}\equiv {1\over 2}\,\Bigl(R(-s)-R(+s)\Bigr),~~~~{\cal C}\equiv
{1\over 2}\,\Bigl(R(-s)+R(+s)\Bigr),~~~{\cal T}\equiv {{\cal
S}\over{\cal C}}.
\end{equation}
Then the STZ equations of motion become:
\begin{equation}
\label{doteps}
\tau_0\,\dot\epsilon^{pl}=\epsilon_0\,{\cal
C}(s)\,\Bigl(\Lambda\,{\cal T}(s)-\Delta\Bigr);
\end{equation}
\begin{equation}
\label{dotdelta}
\tau_0\,\dot\Delta=2\,{\cal C}(s)\,\Bigl(\Lambda\,{\cal
T}(s)-\Delta\Bigr)-\Gamma(s,\Lambda,\Delta)\,\Delta;
\end{equation}
and
\begin{equation}
\label{dotlambda}
\tau_0\,\dot\Lambda=\Gamma(s,\Lambda,\Delta)\,\Bigl(1-\Lambda \Bigr).
\end{equation}
Here, $\epsilon_0 \equiv\lambda\,n_{\infty}$ is roughly the fraction
of the total volume of the low-temperature system in steady-state flow
that is covered by the STZ's.  This is a material-specific quantity.
If $\epsilon_0$ is small, then the disorder induced in the system by
deformation is small.  Conversely, if $\epsilon_0$ is large, then the
STZ-like defects cover the system and the material in some sense
``melts'' under persistent straining.

Throughout this paper, we shall use only what we call the
``quasilinear'' version of these equations.\cite{FLMRS}  That is, we
note that ${\cal T}(s)$ and ${\cal C}(s)$ are, respectively,
antisymmetric and symmetric dimensionless functions of $s$, and write:
\begin{equation}
\label{quasilinear}
{\cal T}(s)\cong {s\over s_y}\equiv \tilde s;~~~~{\cal C}(s)\cong 1,
\end{equation}
where $s_y$ will turn out to be the yield stress.  The choice ${\cal
C}(s)\cong 1$ is, in effect, our definition of the time constant
$\tau_0$.  As pointed out in earlier papers\cite{STZdynsI,FLMRS}, this
quasilinear approximation has important shortcomings.  Neglecting the
stress dependence of ${\cal C}(s)$ means that we overestimate the
amount of plastic deformation that occurs at small stresses and
therefore also overestimate the rate at which orientational memory
disappears in unloaded systems.  Moreover, the quasilinear
approximation is too simplistic to be related directly to atomic
mechanisms, a point that we shall comment upon further in Section V.
Nevertheless, the quasilinear theory has the great advantage that it
is mathematically tractable and easy to interpret. It will serve to
illustrate the main points that we wish to make in this paper, but
aspects of the nonlinearities associated with ${\cal C}$ and ${\cal
T}$ will need to be reintroduced before we shall be able to understand
fully the nonequilibrium behavior of amorphous solids.

Eqs.(\ref{doteps} -\ref{dotlambda}) now become:
\begin{equation}
\label{dotepsql}
\tau_0\,\dot\epsilon^{pl}=\epsilon_0\,(\Lambda\,\tilde s-\Delta);
\end{equation}
\begin{equation}
\label{dotdeltaql}
\tau_0\,\dot\Delta=2\,(\Lambda\,\tilde s-\Delta)-\Gamma(\tilde
s,\Lambda,\Delta)\,\Delta;
\end{equation}
and
\begin{equation}
\label{dotlambdaql}
\tau_0\,\dot\Lambda=\Gamma(\tilde s,\Lambda,\Delta)\,\Bigl(1-\Lambda
\Bigr).
\end{equation}

The quantity $n_{\infty}\,\Gamma/\tau_0$ is the STZ creation rate and
therefore plays exactly the same role as the phenomenological defect
production rates that we discussed -- and complained about -- in the
Introduction. We can derive an expression for that rate by using the
energy-balance argument introduced in \cite{STZdynsI}.  As before, we
start by writing the first law of thermodynamics in the form:
\begin{equation}
\label{energybalance}
2\,\dot\epsilon^{pl}\,s= {2\,
\epsilon_0\,s_y\over\tau_0}\,(\Lambda\,\tilde s-\Delta)\,\tilde s=
\epsilon_0\,s_y\,{d\over dt}\,\psi(\Lambda,\Delta) + {\cal Q}(\tilde
s,\Lambda,\Delta).
\end{equation}
The left-hand side of Eq.(\ref{energybalance}) is the rate at which
plastic work is done by the applied stress $s=s_y\,\tilde s$.  On the
right-hand side, $\epsilon_0\,s_y\,\psi$ is the state-dependent
recoverable internal energy, and ${\cal Q}$ is the dissipation rate.
So long as the STZ's remain uncoupled from the heat bath, ${\cal Q}$
must be positive in order for the system to satisfy the second law of
thermodynamics, that is, for the work done in going around a closed
cycle in the space of variables $s$, $\Lambda$, and $\Delta$ to be
non-negative.

As argued in \cite{STZdynsI}, the simplest and most natural choice for
$\Gamma$ -- and, so far as we can tell, the only one that produces a
sensible theory -- is that it be the energy dissipation rate per STZ.
That is,
\begin{equation}
\label{QGamma}
{\cal Q}(\tilde s,\Lambda,\Delta)=
{\epsilon_0\,s_y\over\tau_0}\,\Lambda\,\Gamma(\tilde s,\Lambda,\Delta),
\end{equation}
With this hypothesis, we can use Eqs. (\ref{dotdeltaql}) and
(\ref{dotlambdaql}) to write Eq.(\ref{energybalance}) in the form
\begin{equation}
\label{energybalance2}
2\,(\Lambda\,\tilde s-\Delta)\,\tilde s=
{\partial\psi\over\partial\Lambda}\, \Gamma\,(1-\Lambda) +
{\partial\psi\over\partial\Delta}\,\Bigl(2(\Lambda\,\tilde
s-\Delta)-\Gamma\,\Delta\Bigr)+\Lambda\,\Gamma.
\end{equation}
Then, solving for $\Gamma$, we find:
\begin{equation}
\label{Gamma1}
\Gamma={2\,(\Lambda\,\tilde s-\Delta)\,(\tilde
s-\partial\psi/\partial\Delta)\over
\Lambda+(1-\Lambda)(\partial\psi/\partial\Lambda)-\Delta\,
(\partial\psi/\partial\Delta)}.
\end{equation}
To assure that $\Gamma$ remains non-negative for all $\tilde s$, we
must let
\begin{equation}
\label{psidelta}
{\partial\psi\over\partial\Delta}= {\Delta\over\Lambda},
\end{equation}
so that the numerator becomes $2\,\Lambda\,(\tilde
s-\Delta/\Lambda)^2$.  Then (see \cite{STZdynsI}), we choose
\begin{equation}
\label{psi}
\psi(\Lambda,\Delta)={\Lambda\over 2}\,\left(1+{\Delta^2\over
\Lambda^2}\right),
\end{equation}
so that
\begin{equation}
\label{Gamma2}
\Gamma(\tilde s,\Lambda,\Delta)={4\,\Lambda\,(\Lambda\,\tilde
s-\Delta)^2\over (1+\Lambda)\,(\Lambda^2-\Delta^2)}.
\end{equation}
This result has the physically appealing feature that it diverges when
$\Delta^2$ approaches its upper limit $\Lambda^2$, thus enforcing a
natural boundary for dynamical trajectories in the space of the state
variables $\Lambda$ and $\Delta$.

It is convenient at this point to replace the variable $\Delta$ by $m
= \Delta/\Lambda$, so that the equations of motion become:
\begin{equation}
\label{doteps2}
\tau_0\,\dot\epsilon^{pl}=\epsilon_0\,\Lambda\,(\tilde s-m);
\end{equation}
\begin{equation}
\label{dotm2}
\tau_0\,\dot m =2\,(\tilde s-m)\,\left[1-{2\,m\,(\tilde s-m)\over
(1+\Lambda)\,(1-m^2)}\right];
\end{equation}
and
\begin{equation}
\label{dotlambda2}
\tau_0\,{\dot\Lambda\over\Lambda}=\left[{4(\tilde s-m)^2\over
1-m^2}\right]\,\left( {1-\Lambda\over 1+\Lambda}\right).
\end{equation}
At the stable fixed point of Eq.(\ref{dotlambda2}), $\Lambda=1$,
Eq.(\ref{dotm2}) becomes
\begin{equation}
\tau_0\,\dot m ={2\,(\tilde s-m)\,(1-\tilde s\,m)\over (1-m^2)},
\end{equation}
which exhibits explicitly the exchange of stability at $\tilde s = 1$
between jammed states with $m=\tilde s$ and flowing states with
$m=1/\tilde s$.

\section{Thermal Effects}

We return now to Eq.({\ref{ndot}), the low-temperature master equation
for the STZ population densities $n_{\pm}$, and ask what changes need
to be made in order to incorporate thermal effects at temperatures
near the glass transition. One obvious possibility is to modify the
rate factors $R(\pm s)$ to include thermal activation across energy
barriers; indeed, we eventually shall have to do that.  (See Section
V.)  However, our quasilinear approximation makes it difficult to do
this systematically.

The more important thermal effects are those that are completely
missing in Eq.({\ref{ndot}), specifically, the thermally assisted
relaxation -- i.e. aging -- of the STZ variables that can occur
spontaneously in the absence of external driving or plastic strain
rate.  There are two ways in which relaxation must occur.  First,
thermal fluctuations ought to act much like deformation-induced
disorder in causing the $n_{\pm}$ to relax toward their steady-state
values $n_{\infty}/2$.  Second, there should be some annealing
mechanism that causes the total STZ population to decrease.  Both of
these mechanisms involve dilations and contractions of the kind
associated with creation and annihilation of STZ's; thus, again for
simplicity, we assume that there is just a single relaxation rate,
denoted $\rho(T)/\tau_0$, that characterizes them.  As we shall see,
that rate will have the Vogel-Fulcher or Cohen-Grest\cite{COHEN-GREST}
form, rapidly becoming extremely small as the temperature $T$ falls
below the glass temperature.  Specifically we expect that $\rho(T)$
has the form
\begin{equation}
\label{rhoT}
\rho(T)=\rho_0\,\exp\left(-{\Delta V^{dil}\over v_f(T)}\right),
\end{equation}
where $\rho_0$ is a dimensionless prefactor, $\Delta V^{dil}$ is the
activation volume required to nucleate a dilational rearrangement, and
$v_f(T)$ is usually identified as the free volume.  The Cohen-Grest
approximation for $v_f(T)$ has the form
\begin{equation}
\label{vfCG}
{v_f(T)\over v_0}=T-T_0 + \sqrt{(T-T_0)^2 + T_1\,T },
\end{equation}
where $v_0$, $T_0$ and $T_1$ are fitting parameters.  This expression
was found by Masuhr et al. \cite{MASUHR} and Lu et al. \cite{LU} to
provide a fairly accurate fit to their data. (We shall 
not use this formula explicitly in what follows.)

In accord with the preceding remarks, our proposed form for the
modified master equation is:
\begin{eqnarray}
\label{ndot2}
\nonumber \tau_0\,\dot n_{\pm}&=&R(\mp s)\,n_{\mp}-R(\pm
s)\,\,n_{\pm}\cr\\ &+&
\Bigl(\Gamma(s,\Lambda,\Delta)+\rho(T)\Bigr)\,\left({n_{\infty}\over
2}-n_{\pm}\right)- \kappa\,\rho(T)\,\left({n_++n_-\over
n_{\infty}}\right)\,n_{\pm}.
\end{eqnarray}
The first and second appearances of $\rho(T)$  on the right-hand side
of Eq.(\ref{ndot2}) correspond, respectively, to its two roles
described above.  The second of these terms, the quadratic form with a
dimensionless multiplicative constant $\kappa$, is similar to the
$n^2$ term on the right-hand side of Eq.(\ref{ndotintro}).   This
bimolecular mechanism has been discussed extensively in references
\cite{TAUB,GREER,TSAO}.

Equations (\ref{dotepsql}-\ref{dotlambdaql}) now become:
\begin{equation}
\label{doteps3}
\tau_0\,\dot\epsilon^{pl}=\epsilon_0\,(\Lambda\,\tilde s-\Delta),
\end{equation}
(unchanged from before);
\begin{equation}
\label{dotdelta3}
\tau_0\,\dot\Delta=2\,(\Lambda\,\tilde s-\Delta)-\Bigl(\Gamma(\tilde
s,\Lambda,\Delta)+\rho(T)\Bigr)\,\Delta -
\kappa\,\rho(T)\,\Lambda\,\Delta;
\end{equation}
and
\begin{equation}
\label{dotlambda3}
\tau_0\,\dot\Lambda=\Bigl(\Gamma(\tilde
s,\Lambda,\Delta)+\rho(T)\Bigr)\,(1-\Lambda)-\kappa\,\rho(T)\,\Lambda^2.
\end{equation}

The next step is to repeat the energy-balance analysis of
Eqs.(\ref{energybalance} - \ref{Gamma2}) to recompute the function
$\Gamma(\tilde s,\Lambda,\Delta)$.  We assert that Eq.(\ref{QGamma})
relating $\Gamma$ and the dissipation rate ${\cal Q}$ must remain
unchanged by the addition of the thermal relaxation terms in
Eq.(\ref{ndot2}).  That is, $\Gamma$ in Eq.(\ref{ndot2}) must be the
energy dissipated per STZ when plastic work is done on the system.
The expression for the internal energy $\psi(\Lambda,\Delta)$ must
remain as given by Eq.(\ref{psi}) because Eq.(\ref{psidelta}) is still
required by the non-negativity condition.  The result, after inserting
the terms proportional to $\rho(T)$ into Eqs.(\ref{energybalance2})
and (\ref{Gamma1}), and transforming to $m=\Delta/\Lambda$, is:
\begin{eqnarray}
\label{tildeGamma}
\nonumber \Gamma(\tilde s,\Lambda,m)+\rho(T) &=&
\Lambda\,\left[{4\,(\tilde s-m)^2 + 2\,\rho(T)
+\kappa\,\rho(T)\,\Lambda\,(1+m^2) \over
(1+\Lambda)\,(1-m^2)}\right]\cr\\&\equiv& \Lambda\,\tilde\Gamma(\tilde
s,\Lambda,m,T).
\end{eqnarray}
Note that $\tilde\Gamma$ is non-negative, as is necessary  because it
is a prefactor for the annihilation and creation rates.  The
non-negativity condition no longer strictly applies to $\Gamma$ itself
because the system is now coupled to the heat bath.

The new equations of motion for $m$ and $\Lambda$ are:
\begin{equation}
\label{doteps4}
\tau_0\,\dot\epsilon^{pl}=\epsilon_0\,\Lambda\,(\tilde s-m),
\end{equation}
\begin{equation}
\label{dotm4}
\tau_0\,\dot m = 2\,(\tilde s -m)-m\,\tilde\Gamma(\tilde
s,\Lambda,m,T),
\end{equation}
and
\begin{equation}
\label{dotlambda4}
\tau_0\,{\dot\Lambda\over\Lambda}= \tilde\Gamma(\tilde
s,\Lambda,m,T)\,(1-\Lambda )-\kappa\,\rho(T)\,\Lambda.
\end{equation}

As in previous presentations, we recover Bingham-like plasticity for
small $\rho$ and for stresses above $s_y$, i.e. $\tilde s > 1$.  In
that case, the equations of motion revert to
Eqs.(\ref{doteps2}-\ref{dotlambda2}) so that, in steady state,
$\Lambda\to 1$ and $m\to 1/\tilde s$.  Thus,
\begin{equation}
\label{bingham}
\dot\epsilon^{pl}\approx {\epsilon_0\over\tau_0}\,\left(\tilde s
-{1\over\tilde s}\right);~~~~~\tilde s > 1.
\end{equation}
This theory exhibits no power-law rheology for flow above the yield
stress.  As we shall see, however, it does exhibit behavior that looks
like superplasticity.

The more interesting behavior is thermally assisted creep below the
yield stress, $\tilde s < 1$, and the transition between creep and
flow near $\tilde s=1$.  For present purposes, we only need to consider
cases in which $\rho(T)\ll 1$.  In the steady-state creep region
where $\tilde s\ll 1$ and $m\cong \tilde s$, Eqs.(\ref{tildeGamma})
and (\ref{dotm4}) tell us that $\tilde s-m$ is small of order $m \rho$.
Then the quantity $(\tilde s-m)^2$ in the numerator of $\tilde\Gamma$
is negligible compared to the other terms.  The steady-state
version of (\ref{dotlambda4}) becomes simply
\begin{equation}
1-\Lambda_N(\kappa) \approx \kappa\,\Lambda_N^2(\kappa);
~~~~\Lambda_N(\kappa) = {1\over 2\,\kappa}\,\left(\sqrt{1+4\,\kappa} -
1\right),
\end{equation}
where the subscript $N$ denotes the low-stress Newtonian limit.

In the complete absence of a driving stress $\tilde s$, the exact
steady-state solutions of Eqs.(\ref{dotm4}) and (\ref{dotlambda4}) are
$m=0$ and $\Lambda=\Lambda_N(\kappa)$.  Thus, the analog of $n_{eq}$
in Eq.(\ref{ndotintro}) is $n_{\infty}\,\Lambda_N(\kappa)$, which is a
temperature-independent quantity instead of being, as is usually
assumed, a rapidly varying function of the form $\exp(-\Delta
V^{dil}/v_f(T))$.  Indeed, the temperature dependence of $n_{eq}$ is
often invoked in the context of equations such as Eq.(\ref{ndotintro})
to interpret calorimetric data.\cite{DEHEY,DUINE}  Because of its lack
of a temperature-dependent $n_{\infty}$, our present theory cannot
predict the specific heat peak near the glass transition that is seen
by the latter authors.  We could fix this problem in an {\it ad hoc}
manner by assigning some temperature dependence to $n_{\infty}$ and/or
$\kappa$.  However, such a procedure would simply gloss over the
fundamental difficulty that we are facing here -- that the limiting
steady-state value of $\Lambda$ must depend upon the order in which we
take the limits $T\to 0$ and $\dot\epsilon^{pl}\to 0$, but that no
such behavior appears in our equations.  This is the same situation
that we discussed in the paragraphs following our first introduction
of $n_{\infty}$ in Eq.(\ref{ndot}); we shall return to it in Section
V.  For the present, we leave the situation as is, with a
temperature-independent $\Lambda_N$, and with the understanding that
we cannot yet use these equations to describe behavior much above the
glass temperature.

To compute the viscosity, it is easiest first to set $\dot m =
\dot\Lambda = 0$ in Eqs.(\ref{dotm4}) and (\ref{dotlambda4}) and
eliminate $\tilde\Gamma$ to find
\begin{equation}
(\tilde s-m)\,(1-\Lambda)={\kappa\,\rho(T)\over 2}\, m\,\Lambda,
\end{equation}
which is an exact relationship between the steady-state values of
$\tilde s$, $\Lambda$ and $m$ at any strain rate.  Combining these
results in the limit $\tilde s \approx m \to 0$, we compute the
Newtonian viscosity $\eta_N$:
\begin{equation}
\label{etaN}
\eta_N \equiv \lim_{\dot\epsilon^{pl}\to 0}\,{s\over
{2\, \dot\epsilon^{pl}}} = {s_y\,\tau_0\over\epsilon_0\,\rho(T)},
\end{equation}
which confirms our expectation that $\rho(T)$ is the rate function
that governs viscous relaxation.

\section{Analysis and Comparison with Experiments}

\begin{figure}[t]
      \includegraphics[angle=-90,
      width=0.6\columnwidth]{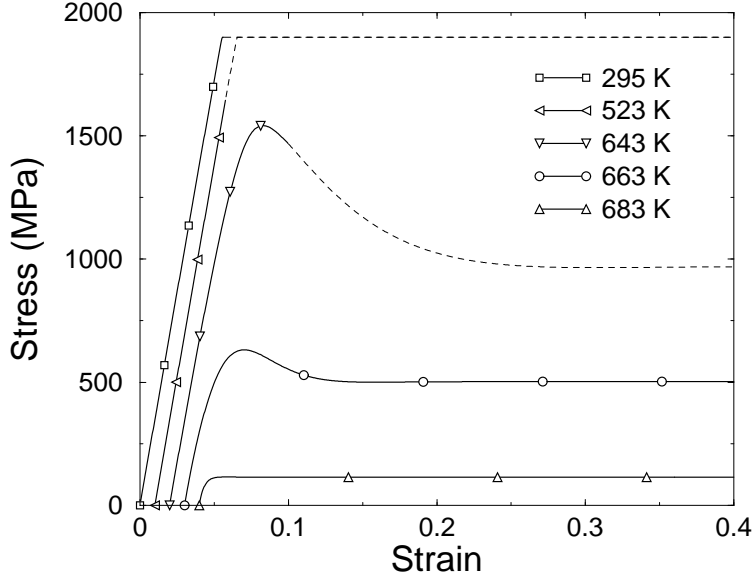} 
      %Other options for include graphics are: 
      %  bb=a b c d, height=h, clip=true, angle=a (and more) 
      \caption{Theoretical curves of tensile stress versus strain for the bulk 
metallic glass ${\rm Zr_{41.2}\,Ti_{13.8}\,Cu_{12.5}\,Ni_{10}\,Be_{22.5}}$ at 
several different temperatures as shown. The strain rate is 
$\dot{\epsilon}^{total} = 1 \times 10^{-1} s^{-1}$. For clarity, the curves 
have been displaced by constant increments along the strain axis.} \label{lufig1}
\end{figure}

Before making comparisons with experiments, we must return to the question of 
how to generalize our two-dimensional theory into one that can be applied to 
three-dimensional experiments. The basic structure of our equations of motion 
must be preserved, with due respect for the relevant symmetries, in any 
generalization of this theory to higher dimensions.  That is, our variables 
$\tilde s$, $\dot\epsilon^{pl}$ and $m$ must become traceless symmetric 
tensors, and $\Lambda$ must remain a scalar, so that Eqs.(\ref{doteps4}) 
and (\ref{dotm4}) become:  
\begin{equation}
\label{doteps5}
\tau_0\,\dot\epsilon_{ij}^{pl}=\epsilon_0\,\Lambda\,(\tilde s_{ij}-m_{ij}),
\end{equation}
and
\begin{equation}
\label{dotm5}
\tau_0\,\dot m_{ij} = 2\,(\tilde s_{ij} -m_{ij})-\tilde\Gamma(\tilde
s,\Lambda,m,T)\,m_{ij}.
\end{equation}
The energy-balance analysis yields
\begin{equation}
\label{tildeGammatensor}
\tilde\Gamma(\tilde s,\Lambda,m,T)= {2\,(\tilde s_{ij}-m_{ij})\,(\tilde s_{ij}-m_{ij}) + 2\,\rho(T)+\kappa\,\rho(T)\,\Lambda\,(1+\bar m^2) \over
(1+\Lambda)\,(1-\bar m^2)},
\end{equation}
where we are using the summation convention, and $\bar m^2 = (1/2)\,m_{ij}\,m_{ij}$.  

The experiments in which we are interested involve only uniaxial
stresses, say, in the $x$ direction.  If there are no tractions in the
$y$ or $z$ directions, and if we can ignore spatial nonuniformities,
then the total stress $\sigma_{ij}$ is diagonal with
$\sigma_{xx}=\sigma$, $\sigma_{yy}=\sigma_{zz}=0$.  Similarly, for the
deviatoric stress, $\tilde s_{yy}=\tilde s_{zz}=- \tilde s_{xx}/2$;
and $m_{yy}=m_{zz}=-m_{xx}/2$. Thus, $\bar m^2 = 3/4\,m_{xx}^2$. We
can now make the $xx$ components of Eqs. (\ref{doteps5})--(\ref{tildeGammatensor})
 look exactly like Eqs. (\ref{doteps4}),
(\ref{dotm4}), and (\ref{tildeGamma}) by defining $m^2 = \bar m^2$,
$m=\sqrt{3/4}\,m_{xx}$, $\tilde s = \sqrt{3/4}\,\tilde s_{xx}$,  
$\dot\epsilon^{pl} = \dot\epsilon_{xx}^{pl}$, and by
replacing $\epsilon_0$ by $\epsilon_0' = \sqrt{4/3}\,\epsilon_0$.
Note that the dynamical exchange of stability, i.e. plastic yielding,
still occurs at $\tilde s= 1$; thus our scaling of the stress by $s_y$
remains correct.   When comparing to the data, which is presented in
terms of the uniaxial stress $\sigma=\sigma_{xx}$,  we write
$\sigma=(3/2) s_{xx} = \sqrt{3} s_y\,\tilde s$ or, equivalently,
$\sigma = \sigma_y\,\tilde s$, where $\sigma_y$ is the tensile yield
stress.  Equation (\ref{etaN}), however, remains unchanged with
$\epsilon_0$ rather than $\epsilon_0'$,  because viscosity now is
expressed as $\eta = s_{xx}/ 2 \dot\epsilon^{pl}_{xx}=
s /  \sqrt{3} \dot\epsilon^{pl}$.

To illustrate the principal results of our analysis, we first follow
the lead of Kato et al \cite{KATO} and Lu et al \cite{LU} by looking
for scaling in the steady-state behavior of our system.  To be
specific about what we mean here, we show in Figs. 1 and 2 sets of
stress-strain curves for various temperatures and fixed strain rates.
As we shall explain shortly, these figures are to be compared with
Figs. 1 and 2 in \cite{LU}.  (See also Fig. 1 in \cite{DEHEY} and Fig. 1
in \cite{KATO}.)  A general feature of these curves is that, when the
strain rate is held constant, the stress rises through a maximum,
decreases as the material softens, and then reaches a steady-state
value.  We shall discuss the initial transients later in this Section,
but look first at the late-stage, steady-state behavior.

\begin{figure}[t]
      \includegraphics[angle=-90,
      width=0.6\columnwidth]{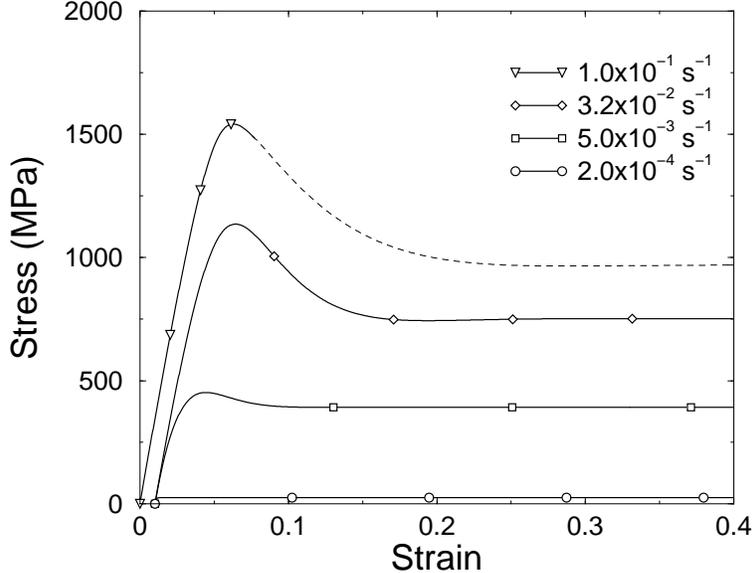} 
      %Other options for include graphics are: 
      %  bb=a b c d, height=h, clip=true, angle=a (and more) 
       \caption{Theoretical curves of tensile stress versus strain for 
the bulk metallic glass 
${\rm Zr_{41.2}\,Ti_{13.8}\,Cu_{12.5}\,Ni_{10}\,Be_{22.5}}$ at 
several different strain rates as shown. The temperature 
is $T = 643\,$K. For clarity, all but the first of these curves have been 
displaced by the same amount along the strain axis.} \label{lufig2}
\end{figure}

We compute the steady-state flow stress as a function of the strain
rate by solving Eqs.(\ref{dotm4}) and (\ref{dotlambda4}) with $\dot m
= \dot\Lambda=0$.  Then, as in \cite{KATO} and \cite {LU}, we plot
$\tilde s = s/s_y$ as a function of $\eta_N\,\dot\epsilon^{pl}$ for
eight  different values of the relaxation rate $\rho(T)$ corresponding
to the eight different temperatures for which data are reported in
\cite{LU}.  The results are shown in Fig. 3.  As discovered by Kato et
al \cite{KATO}, all of these curves lie on top of one another for
stresses $\tilde s < 1$ but, in our case, they diverge from each other
in the flowing regime, $\tilde s > 1$, where the Bingham-like behavior
shown in Eq.(\ref{bingham}) sets in.

Figure (4a) contains the same theoretical curves as those shown in
Fig. 3, but plotted there as tensile stress versus scaled strain rate,
and compared with experimental data taken from Fig. 9a of \cite{LU}.
The same theoretical functions and data points are replotted in
Fig. 4b to show the normalized viscosity, $\eta/\eta_N$ as a function
of the scaled strain rate. The latter figure is directly comparable to
\cite{LU}, Fig. 9b.  Note that the range of strain rates shown in
Figs. 4 corresponds to the range of the experimental data and is
substantially smaller than that shown in Fig. 3.  The theoretical
curve that lies above the rest at high strain rates is for $T=683\,K$,
the highest of the temperatures reported in \cite{LU}.  The data
points at that temperature all lie at scaled strain rates that are too
small to test this predicted breakdown of the scaling law.

Our Fig. 5a shows individual theoretical and experimental curves of
tensile stress as a function of (unscaled) strain rate for different
temperatures.  Here, the experimental data is from \cite{LU}, Fig. 7.
These curves are replotted in Fig. 5b to show (unscaled) viscosity as
a function of strain rate, analogous to \cite{LU}, Fig. 8.

\begin{figure}[t]
      \includegraphics[angle=-90,
      width=0.6\columnwidth]{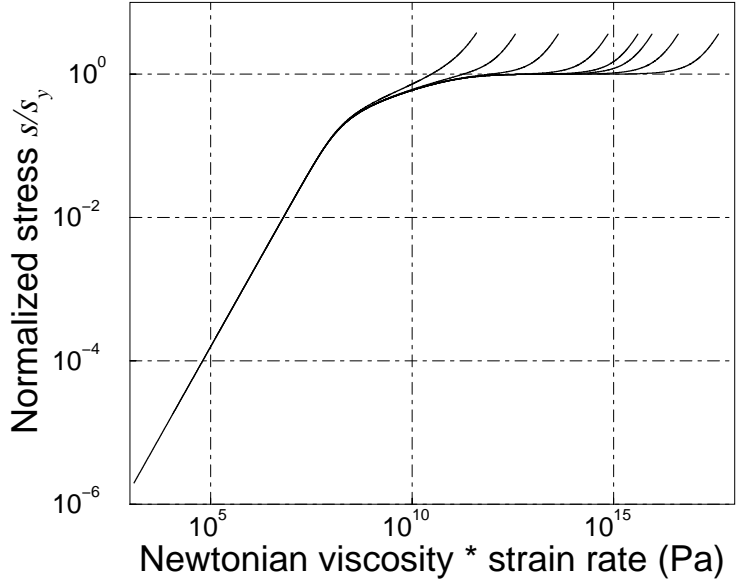} 
      %Other options for include graphics are: 
      %  bb=a b c d, height=h, clip=true, angle=a (and more) 
\caption{Scaling behavior in the STZ theory: shear stress $\tilde{s}$ as 
a function of strain rate scaled by $\eta_N$. This graph is plotted for 
the same set of temperatures as shown in Fig. \ref{fig9ab}a, 
but for a larger range of strain rates.}  \label{fig3}
\end{figure}

\begin{figure}[t]
      \includegraphics[angle=-90,
      width=0.5\columnwidth]{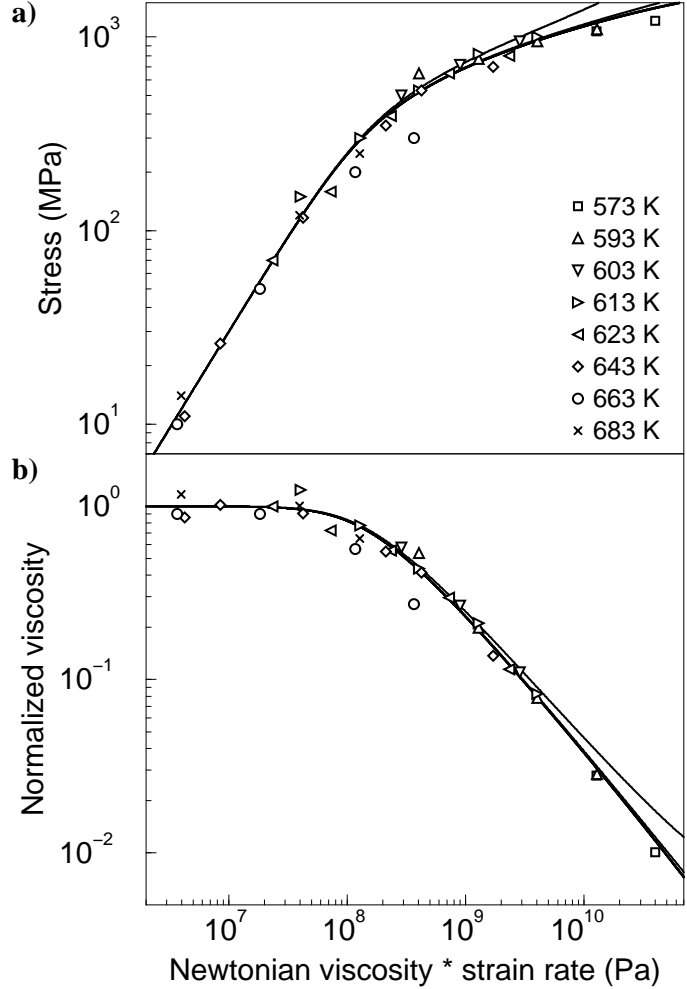} 
      %Other options for include graphics are: 
      %  bb=a b c d, height=h, clip=true, angle=a (and more) 
      \caption{Tensile stress and viscosity as functions of scaled 
strain rate $\eta_N \dot{\epsilon}^{total}$.  The data points for the 
bulk metallic glass 
${\rm Zr_{41.2}\,Ti_{13.8}\,Cu_{12.5}\,Ni_{10}\,Be_{22.5}}$ are taken 
from \cite{LU}, Figs. 9a and 9b.  The solid curves are theoretical 
results computed for the same set of temperatures as shown.} \label{fig9ab}
\end{figure}

In constructing these figures, we have determined our theoretical
parameters as follows:
\begin{itemize}

\item[]We have used the value of the room-temperature tensile yield stress
reported in \cite{LU}, $\sigma_y=1.9$ GPa.  Thus $\sigma= 1.9\,\tilde s$ GPa.

\item[]Rather than using the Cohen-Grest formula with parameters from
\cite{MASUHR}, we have taken the limiting (vanishing strain rate)
Newtonian viscosities directly from \cite{LU}, Fig. 10, and have checked
that these values (apart from one apparently misplaced point) are
consistent with the data points in \cite{LU}, Fig. 9.

\begin{figure}[t]
      \includegraphics[angle=-90, width=0.5\columnwidth]{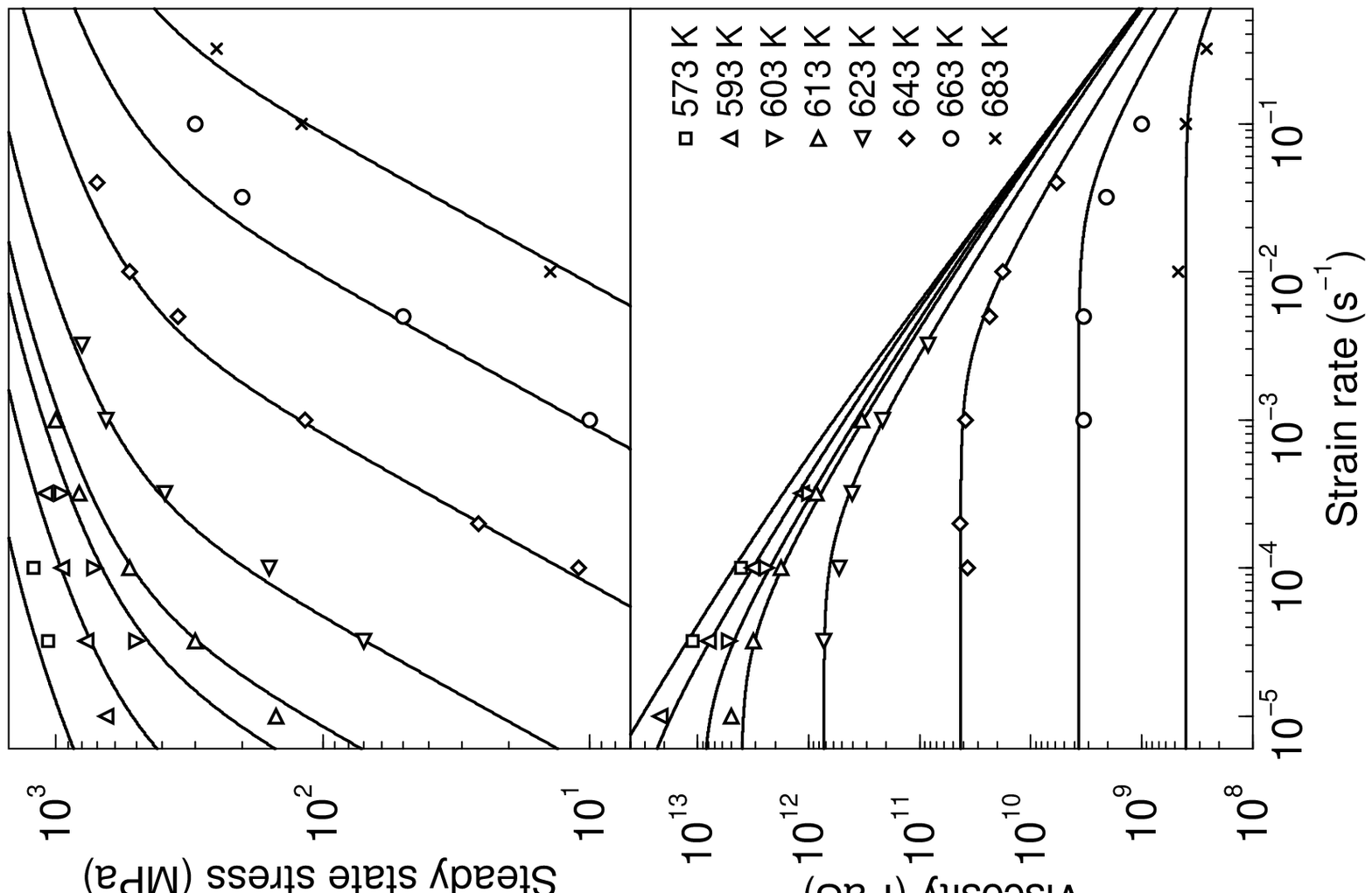}
      %Other options for include graphics are: 
      %  bb=a b c d, height=h, clip=true, angle=a (and more) 
\caption{Tensile stress (a) and viscosity (b) as functions of strain 
rate for different temperatures as shown. The data points are for the 
bulk metallic glass 
${\rm Zr_{41.2}\,Ti_{13.8}\,Cu_{12.5}\,Ni_{10}\,Be_{22.5}}$ 
as reported by \cite{LU},  Figs. 7 and 8.  The solid lines 
are theoretical curves.} \label{fig78}
\end{figure}

\item[]Given the above constraints, we are left with only two parameters,
$\kappa$ and $\epsilon_0/\tau_0$ (or, equivalently, $\epsilon_0'$), that can be adjusted to fit the
steady-state experimental data in \cite{LU}, Fig. 7.  Because we know
experimental values for the limiting viscosities $\eta_N(T)$ at
different temperatures, a value for the ratio $\epsilon_0/\tau_0$ in
Eq.(\ref{etaN}) determines the overall scale factor for the function
$\rho(T)$.  Thus our fitting procedure has been  to start by choosing
values of $\kappa$ and $\epsilon_0/\tau_0$.  We then use those  values
to determine $\rho(T)$ and to compute steady-state solutions of
Eqs.(\ref{dotm4}), (\ref{dotlambda4}) and (\ref{doteps2}).  From these
solutions we compute steady-state stress versus strain-rate relations
that can be compared with the experimental data.  We then iterate this
process to find best-fit values for $\kappa$ and $\epsilon_0/\tau_0$.

\end{itemize}

Our best-fit parameter values, obtained by the procedure outlined
above, are  $\kappa=120$ and $\epsilon_0/\tau_0 = 260\,s^{-1}$.  Our
corresponding values of $\rho(T)$, along with our estimates of the
viscosities, are shown in Table 1.  We emphasize that these values are
not much better than order-of-magnitude estimates.  
  
As is obvious in Fig. 5, there is scatter in the experimental data,
and there may also be systematic errors.  For example, the two  points
at the highest strain rates for $T = 663\,$K fall well below our
theoretical curve for that temperature while the theoretical fits look
good for the temperatures on either side, i.e. $T = 643\,$K and
$683\,$K.  We  could improve the fits to all three of these curves by
choosing a  substantially larger value of $\kappa$; but we would do
this at the  expense of poorer fits at lower temperatures. However,
the  values of the viscosities that we can deduce from
\cite{LU}, Fig. 10 seem  uncertain, possibly by factors of 2 or 3; so
our estimates of $\rho(T)$  and, therefore, all our theoretical curves
-- especially those at the  lower temperatures -- might be modified by
more accurate viscosity data.

Within the above uncertainties, our theoretical fits to the
experimental  data are relatively insensitive to our choices of the
two parameters that we have allowed ourselves.  On the one hand, this
insensitivity gives us confidence in the basic structure of our
theory; on the other hand, it means that we cannot yet test the theory
in as much detail as we would like.  For example, the yield stress
$s_y$ probably ought to be a function of temperature, decreasing
slowly (in contrast to the rapidly varying viscosity) from room
temperature through the experimental range.  Its behavior should be
roughly like that of the shear modulus, which must soften as the
system approaches the glass transition.  Also, as we have discussed
above,  we expect that $\epsilon_0$ ultimately will be temperature
dependent because it is proportional to $n_{\infty}$.  We could
improve the fits, for example, at the lower temperatures shown in
Fig. 5, by incorporating such temperature dependences into our
equations; but it seems to make little sense to do so without first
resolving various uncertainties in both the theory and the experiments.

\squeezetable
\begin{table}[t]
\begin{ruledtabular}
\begin{tabular}{|c|c|c|c|c|c|c|c|c|}
Temperature, K & 573 & 593 & 603 & 613 & 623 & 643 & 663 & 683 \\
Viscosity, PaS & $4.00 \times10^{14}$  & $4.03\times10^{13}$ & $8.99\times10^{12}$  &  $4.03 \times10^{12}$  &  $7.29\times10^{11}$ &  $4.27\times10^{10}$  & $3.68\times10^{9}$   & $3.99\times10^{8}$  \\
$\rho$         & $1.07\times 10^{-8}$ & $1.06\times 10^{-7}$ & $4.77 \times 
10^{-7}$ & $1.06 \times 10^{-6}$ &  $5.88 \times 10^{-6}$ & 
$ 1.01 \times 10^{-4}$ & $1.17 \times 10^{-3}$ & $1.15 \times 10^{-2}$ \\
\end{tabular}
\end{ruledtabular}\caption{\label{table1}Experimental data for viscosity taken from \cite{LU}, and values of $\rho$ used in the present calculations.}
\end{table}

Our main conclusion from this steady-state analysis is that we are
observing a transition from thermally assisted creep to viscoplastic
flow in the neighborhood of the dynamic yield stress.  At low stresses
and strain rates, the linear response relation contains only the
factor $\eta_N \propto 1/\rho(T)$, thus we obtain the simple scaling.
Near the yield stress, however, our theoretical strain rate increases
by several orders of magnitude for small increments of stress, and the
experimental behavior tracks this trend accurately.  This behavior
resembles superplasticity.  Interestingly, the theoretical scaling
persists through the ``superplastic'' region and does not break down
until true viscoplastic flow begins.

\begin{figure}[t]
      \includegraphics[angle=-90,
      width=0.6\columnwidth]{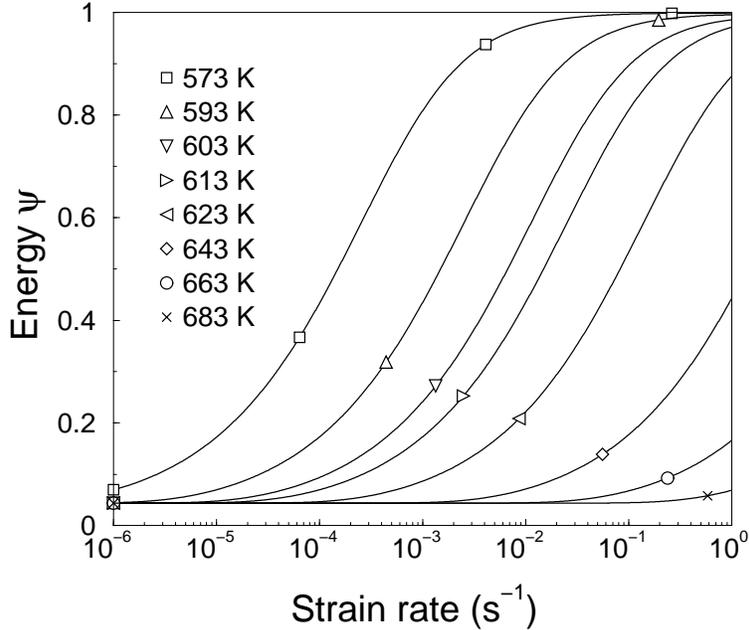} 
      %Other options for include graphics are: 
      %  bb=a b c d, height=h, clip=true, angle=a (and more) 
      \caption{Dimensionless energy $\psi$ as a function of strain rate for
the same set of temperatures shown in the preceding figures.} \label{fig6}
\end{figure}

Before returning to the transient stress-strain curves shown in
Figs. 1 and 2, we show one other steady-state prediction of our theory
for which there is no data in \cite{LU}, but which seems to be
potentially important.  In Fig. \ref{fig6}, we have used our
steady-state solutions for $m$ and $\Lambda$ to plot the dimensionless
internal energy $\psi=(\Lambda/2)\,(1+m^2)$ as a function of the
strain rate $\dot\epsilon^{pl}$, for the set of temperatures used in
the preceding figures.  The  energy $\psi$ might be measured by
differential scanning calorimetry, as has been done by Hasan and Boyce
in studies of polymeric glasses.\cite{HB93,HB95}  De Hey et al
\cite{DEHEY} also report DSC results, which they interpret as
measurements of free volume.   We assume that Fig. 4 in \cite{DEHEY},
apart from the scale on the  vertical axis, is  at least qualitatively
the same as a graph of $\psi$ versus strain rate  analogous to our
Fig. 6. The two figures are similar to one another if we look at our
theoretical curves only at small strain rates.

An important feature of our Fig. 6 is that, at fixed
strain rate, $\psi$ decreases as $T$ increases.  That
trend is exactly what we expect for thermally assisted creep; the
higher the temperature, the fewer STZ's are needed in order to sustain
a given flow rate, and thus the smaller the internal energy.  Note,
however, that our theory predicts that all of these curves converge to
a single, temperature independent, value $\psi\to \Lambda_N(\kappa)/2$
in the limit $\dot\epsilon^{pl}\,\to\,0$.  On the other hand, De Hey
et al. plausibly assert that the equilibrium value of the STZ density,
$n_{eq}$, should be an increasing function of temperature.  If so,
these curves must cross each other and the trend in temperature
dependence must be reversed at small enough values of
$\dot\epsilon^{pl}$. An extension of these measurements to smaller
strain rates might therefore provide a test of our differing
assumptions about the STZ density $n_{eq}$.

So far, we have examined only steady-state behavior.  We turn next to
stress-strain curves obtained in constant strain-rate experiments such
as those shown in our Figs. 1 and 2 and in Figs. 1 and 2 of \cite{LU}.
To plot these curves, we solve
\begin{equation}
\label{tildesdot}
{\sigma_y\over E}\,\dot{\tilde s} = 
\dot{\epsilon}_{xx}^{total}-\sqrt{4\over 3}\,{\epsilon_0\over\tau_0}\,
\Lambda\,(\tilde s - m),
\end{equation}
along with Eqs.(\ref{dotm4}) and (\ref{dotlambda4}) to compute $\tilde
s$ as a function of the total strain.  We use the value of Young's
modulus given in \cite{LU}, $E=96$ GPa, to estimate $ E/\sigma_y \cong
50$. For time dependent calculations, we must choose a value of the
time scale $\tau_0$, which multiplies the time derivatives in
Eqs.(\ref{dotm4}) and (\ref{dotlambda4}).  To do this, we keep the
ratio $\epsilon_0/\tau_0$ fixed at its value obtained from the
steady-state calculations, and we adjust the value of  $\epsilon_0$ so
as to fit the transient stress-strain curves.  Our best-fit value is
$\epsilon_0 =0.7$, which means that $\tau_0\cong 2.7\times
10^{-3}\,s$. We emphasize that this value, like our estimates for
$\kappa$ and $\epsilon_0/\tau_0$, remains highly uncertain.  It is
possible that we may eventually be able to sharpen our estimate of the
time scale $\tau_0$ by using the time-dependent stress-relaxation data
shown in \cite{LU}.  In fact, we can reproduce those results about as
well as the other results shown here, but our theory in this case is
especially sensitive to our quasilinear approximation plus other
uncertainties regarding the experiments.

Our Figs. 1 and 2 are drawn so as to be directly comparable to Figs. 1 and 2 in 
\cite{LU}; that is, we use the same strain rates and temperatures.  
The one other parameter that we must choose for solving Eqs. (\ref{dotm4}),
(\ref{dotlambda4}), and (\ref{tildesdot}) is the initial value of
$\Lambda$, which we denote by $\Lambda_0$.  Theoretically, the
smallest value of $\Lambda_0$ that can be achieved by annealing is
$\Lambda_N(\kappa) = 0.087$ (for $\kappa=120$); therefore we have used this
value of $\Lambda_0$ for these two figures. The initial 
values of $m$ and $\tilde s$ are always chosen to be zero.

In all cases, the agreement between theory and experiments seems
satisfactory given the various uncertainties. The peak heights and
positions for fixed strain rate $\dot\epsilon^{total} = 0.1\,\,
s^{-1}$ and varying temperatures in Fig. 1, and for fixed temperature
$T = 643\,K$ and varying strain rates in Fig. 2, are within about ten
percent of their experimental values.  The experimental curves for low
temperatures and large strain rates end where the samples break; the
dashed lines in our figures indicate our theoretical extensions of
those parts of the curves for which no experimental data is
available. The one systematic discrepancy is that our initial
theoretical slopes are smaller than the experimental ones.  This is an
artifact of our quasilinear approximation, which ignores the strong
stress and temperature dependence of the factor ${\cal C}(s)$ in
Eqs. (\ref{doteps},\ref{dotdelta}).  (See remarks in Section V.) In
the limit of low temperatures and very small strain rates, our theory
predicts this initial slope to be
$E/( 1 + E\,\epsilon_0'\,\Lambda_0/ 2 \sigma_y)$ instead of simply $E$. This
does not happen in the fully nonlinear theory presented in
\cite{FL}. We have chosen not to include these nonlinear effects in
our calculations here because they introduce additional undetermined
parameters, and they are not necessary to describe the shear softening
and shear thinning observed at higher stresses near the glass
transition temperature.

\begin{figure}[t]
      \includegraphics[angle=-90,
      width=0.6\columnwidth]{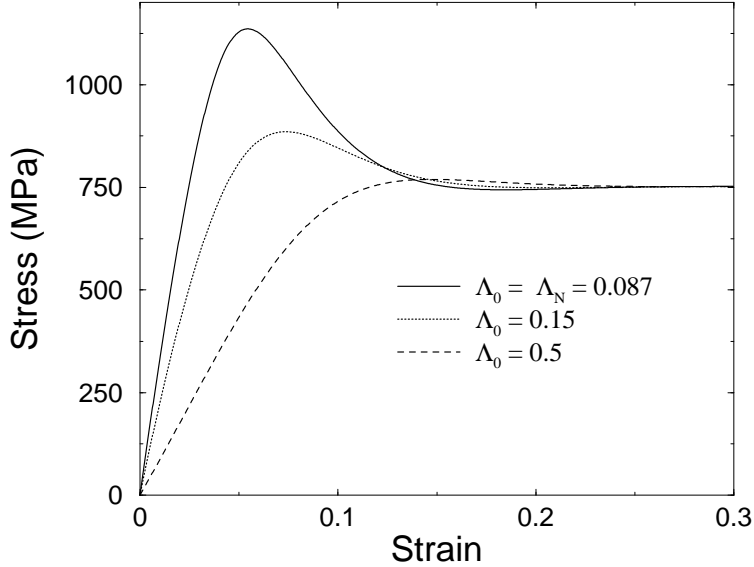} 
      %Other options for include graphics are: 
      %  bb=a b c d, height=h, clip=true, angle=a (and more) 
      \caption{Tensile stress as a function of strain for several different
values of $\Lambda_0$. Curves are plotted for 
$\dot\epsilon^{total}= 3.2 \times 10^{-2}\, s^{-1}$ at $T=643$K.}\label{fig7}
\end{figure}

Finally, in Fig. \ref{fig7}, we use the material parameters deduced above for
the system studied in \cite{LU} to plot stress-strain curves for
different $\Lambda_0$'s, all at temperature $T=643 K$ and
$\dot\epsilon^{total} = 3.2 \times 10^{-2}\, s^{-1}$.  The different $\Lambda_0$'s
correspond to different initial states of disorder produced by varying
the annealing times and temperatures.  Presumably, annealing for
longer times at lower temperatures produces smaller values of
$\Lambda_0$; but it seems difficult to make quantitative estimates of
this effect.  These curves may be compared qualitatively with those
shown in \cite{DEHEY}, Fig. 9, where larger initial densities of STZ's
produce larger plastic responses and correspondingly smaller overshoots during the early stages of deformation.

\section{Concluding Remarks}

The comparison with experiments discussed in the preceding section
leads us to believe that the STZ theory captures the main features of
the experimental data, but that we shall have to improve it in
specific respects if we are to develop it into a yet more
quantitative, predictive description of plastic deformation in
amorphous solids. We conclude this paper by identifying three
directions for the next phases of these investigations.

\noindent {\it Fully nonlinear, temperature dependent transition
rates:} When we examine the quasilinear STZ theory in the context of a
theory that includes thermal fluctuations, we see that it is a special
case in which the shear rearrangements are not being modeled as
realistically as the dilations or contractions.  To see this in more
detail, go back to our original\cite{FL}, fully nonlinear version of the
low-temperature rate factors $R(s)$:
\begin{equation}
\label{Rnonlinear}
R(s) = {1\over\tau_0}\,\exp\left(-{\Delta V^{shear}(s)\over
v_f}\,\right);~~~~\Delta V^{shear}(s)= \Delta
V_0^{shear}\,e^{-s/\bar\mu},
\end{equation}
where $\Delta V^{shear}(s)$ is the activation volume required to
nucleate a shear transformation.  Our idea here was that, at
temperatures well below the glass temperature, the transitions between
STZ states are not thermally activated but, rather, are controlled
entropically.  That is, the rate factors are determined by the number
of paths that the molecules within a zone can follow in moving around
each other while going from one state to the other. The exponential
factor in Eq.(\ref{Rnonlinear}) is an approximation for a weighted measure of that number
of paths.  Its $s$-dependence means that greater weight must be given
to paths moving in the direction of the stress than opposite to it.
The exponential form of $\Delta V^{shear}(s)$ is the simplest
non-negative function that becomes arbitrarily small at large $s$ and
introduces just one new parameter, the effective STZ stiffness
$\bar\mu$.  The quasilinear version of the theory corresponds
(roughly) to the limit of small $s$ and small values of $\Delta
V_0^{shear}/v_f$.

Comparison of Eq.(\ref{Rnonlinear}) with Eq.(\ref{rhoT}) indicates
that the natural way to include thermal effects in $R(s)$ is simply to
let $v_f$ have the $T$-dependent Cohen-Grest form shown in
Eq.(\ref{vfCG}).  This means that, at low $T$, the ratio $\Delta
V_0^{shear}/v_f(T)$ becomes very large,  which, in turn, implies that
the functions ${\cal C}(s)$ and ${\cal T}(s)$ introduced in
Eq.(\ref{Tdef}) become strongly stress dependent, and the quasilinear
approximations made in Eqs.(\ref{quasilinear}) are no longer valid.
Importantly, ${\cal C}(s)$ becomes very small for small $s$, so that
plastic deformation is strongly suppressed at stresses appreciably
below the yield stress.

The strong stress dependence of ${\cal C}(s)$ and ${\cal T}(s)$ should
be especially apparent in transient behavior of the kind shown in
Figures 1, 2 and 7. Here, the initial response to loading at small
stress will be almost entirely elastic, and plastic deformation will
begin only later in the process. We shall have to use the fully
nonlinear theory when undertaking more detailed comparisons with these
kinds of experimental results.

\noindent {\it Shear localization:} All of the analysis in this paper
pertains to spatially homogeneous systems.  In order to make closer
contact with experiments, we shall have to understand why and when
these systems become unstable against shear banding and inhomogeneous
failure modes, especially fracture.

One mechanism for shear localization that we have not mentioned in
this presentation is the elastic interaction between STZ's studied in
\cite{SHEARLOC}.  As shown in that paper, an STZ-like event generates
a quadrupolar stress field that induces other nearby events along
preferred spatial directions and suppresses events elsewhere.  The
result is a tendency toward shear localization that should be
interesting to examine in the context of this more general version of
the STZ theory.

A second mechanism that seems likely to play a role in shear
localization is already built into our equations of motion when we
write them in terms of spatially varying fields.  From
Eqs.(\ref{tildeGamma}) and (\ref{dotlambda4}), we see that the STZ
density $\Lambda$ grows most rapidly, within limits, in regions where
$\Lambda$ already is large.  This feedback effect, perhaps coupled to
the effect of elastic interactions mentioned above, is our best guess
at present about how shear banding will emerge in the STZ theory.

\noindent {\it Effective temperature and the interpretation of
$n_{eq}$:} Finally, we return to Spaepen's suggestion that the density
of STZ's might be directly related to the free volume as in
Eq.(\ref{vfintro}).  Note that $v_f$ appears in that equation, not as
an extensive quantity related directly to the difference between the
actual volume and some hypothetical close-packed volume, but rather as
an intensive variable analogous to a temperature.  For example, Mehta
and Edwards\cite{ME} have introduced an intensive variable,
thermodynamically conjugate to volume, which must govern density
fluctuations in relations such as Eq.(\ref{vfintro}) or (\ref{rhoT})
in much the same way as ordinary temperature governs energy
fluctuations. One of us (MLF)\cite{FALKband,  FALKband1} has shown in a recent
molecular-dynamics simulation that while, as expected, the interior of
a shear band has a high density of active STZ's, the actual free
volume in that region is only slightly greater than in the rest of the
system, but the intensity of density fluctuations -- both compressive
and dilational -- is substantially higher.

The idea that $\chi\equiv v_f/V^*$ in Eq.(\ref{vfintro}) might more
generally be interpreted as an effective temperature seems especially
appealing in light of our argument that the limiting steady-state
value of $\Lambda$ (or equivalently $n_{\infty}$ or $n_{eq}$) should
depend upon the order in which we take the limits $T\to 0$ and
$\dot\epsilon^{pl}\to 0$. More realistically, under steady-state
conditions, $\chi$ might  approach some nonzero limiting value and
remain there for indefinitely long times at sufficiently small $T$ and
for arbitrarily  small but nonzero $\dot\epsilon^{pl}$.  Conversely,
$\chi$ (in suitable units) might approach the true temperature at
large enough $T$ or small enough $\dot\epsilon^{pl}$.

There is increasing evidence that something like this happens in
sheared foams or granular
materials.\cite{Teff,CUGLIANDOLOetal,SOLLICHetal,BERTHIER-BARRAT} In
those systems, the usual kinetic temperature is zero because the
constituents have very large masses, but an effective temperature
determined by response-fluctuation relations goes to a nonzero limit
when the deformation rate becomes arbitrarily small.  In our present
system, there is a true kinetic temperature, but below the glass
transition that temperature is so small that thermally assisted
molecular rearrangements are effectively frozen out.  During
irreversible processes such as plastic deformation, therefore, the
slow, configurational degrees of freedom characterized by $\chi$ might
fall out of equilibrium with the fast, thermal (vibrational) degrees
of freedom, and each may accurately be described by its own
``temperature.'' We suspect that some such description of our system
will be necessary in order to resolve the two-limit problem that we
have encountered here.

\begin{acknowledgments}
J. Langer and L. Pechenik were supported primarily by U.S. Department
of Energy Grant No. DE-FG03-99ER45762, and in part by the MRSEC Program of the National Science Foundation under Award No. DMR96-32716. M. Falk was supported by the National Science Foundation under Award No. DMR-0135009, and in part by the Dow Corning Foundation.
\end{acknowledgments}

\end{document}